\documentclass[preprint, onecolumn,10pt]{article}
\usepackage[super,sort&compress,comma]{natbib} 
\usepackage[version=3]{mhchem}
\usepackage[left=1.5cm, right=1.5cm, top=1.785cm, bottom=2.0cm]{geometry}
\usepackage{balance}
\usepackage{mathptmx}
\usepackage{sectsty}
\usepackage{graphicx} 
\graphicspath{{figures/}}
\usepackage{lastpage}
\usepackage[format=plain,justification=justified,singlelinecheck=false,font={stretch=1.125,small,sf},labelfont=bf,labelsep=space]{caption}
\usepackage{float}
\usepackage{fancyhdr}
\usepackage{fnpos}
\usepackage[english]{babel}
\addto{\captionsenglish}{%
}
\usepackage{array}
\usepackage{droidsans}
\usepackage{charter}
\usepackage[utf8]{inputenc}
\usepackage[T1]{fontenc}
\usepackage[usenames,dvipsnames]{xcolor}
\usepackage{setspace}
\usepackage[compact]{titlesec}
\usepackage{hyperref}

\usepackage{ulem} 
\usepackage{siunitx}
\usepackage{amsmath,amssymb}
\usepackage{epstopdf}

\definecolor{cream}{RGB}{222,217,201}

\begin{document}

\pagestyle{fancy}
\thispagestyle{plain}
\fancypagestyle{plain}{
\renewcommand{\headrulewidth}{0pt}
}

\makeFNbottom
\makeatletter
\renewcommand\LARGE{\@setfontsize\LARGE{15pt}{17}}
\renewcommand\Large{\@setfontsize\Large{12pt}{14}}
\renewcommand\large{\@setfontsize\large{10pt}{12}}
\renewcommand\footnotesize{\@setfontsize\footnotesize{7pt}{10}}
\makeatother

\renewcommand{\thefootnote}{\fnsymbol{footnote}}
\renewcommand\footnoterule{\vspace*{1pt}%
\color{cream}\hrule width 3.5in height 0.4pt \color{black}\vspace*{5pt}} 
\setcounter{secnumdepth}{5}

\makeatletter 
\renewcommand\@biblabel[1]{#1} 
\renewcommand\@makefntext[1]%
{\noindent\makebox[0pt][r]{\@thefnmark\,}#1}
\makeatother 
\renewcommand{\figurename}{\small{Fig.}~}
\sectionfont{\sffamily\Large}
\subsectionfont{\normalsize}
\subsubsectionfont{\bf}
\setstretch{1.125} 
\setlength{\skip\footins}{0.8cm}
\setlength{\footnotesep}{0.25cm}
\setlength{\jot}{10pt}
\titlespacing*{\section}{0pt}{4pt}{4pt}
\titlespacing*{\subsection}{0pt}{15pt}{1pt}

\fancyfoot{}
\fancyfoot[LO,RE]{\vspace{-7.1pt}\includegraphics[height=9pt]{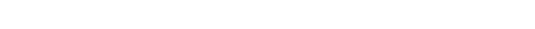}}
\fancyfoot[CO]{\vspace{-7.1pt}\hspace{13.2cm}\includegraphics{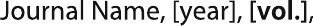}}
\fancyfoot[CE]{\vspace{-7.2pt}\hspace{-14.2cm}\includegraphics{head_foot/RF}}
\fancyfoot[RO]{\footnotesize{\sffamily{1--\pageref{LastPage} ~\textbar \hspace{2pt}\thepage}}}
\fancyfoot[LE]{\footnotesize{\sffamily{\thepage~\textbar\hspace{3.45cm} 1--\pageref{LastPage}}}}
\fancyhead{}
\renewcommand{\headrulewidth}{0pt} 
\renewcommand{\footrulewidth}{0pt}
\setlength{\arrayrulewidth}{1pt}
\setlength{\columnsep}{6.5mm}
\setlength\bibsep{1pt}

\makeatletter 
\newlength{\figrulesep} 
\setlength{\figrulesep}{0.5\textfloatsep} 

\newcommand{\topfigrule}{\vspace*{-1pt}%
\noindent{\color{cream}\rule[-\figrulesep]{\columnwidth}{1.5pt}} }

\newcommand{\botfigrule}{\vspace*{-2pt}%
\noindent{\color{cream}\rule[\figrulesep]{\columnwidth}{1.5pt}} }

\newcommand{\dblfigrule}{\vspace*{-1pt}%
\noindent{\color{cream}\rule[-\figrulesep]{\textwidth}{1.5pt}} }

\makeatother

{\includegraphics[height=30pt]{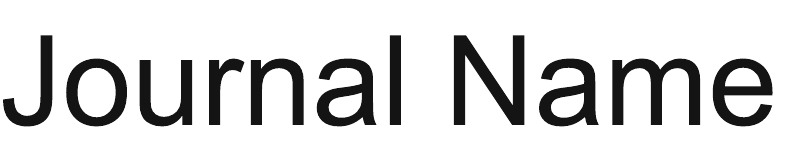}\hfill\raisebox{0pt}[0pt][0pt]  {}\\[1ex]
\includegraphics[width=18.5cm]{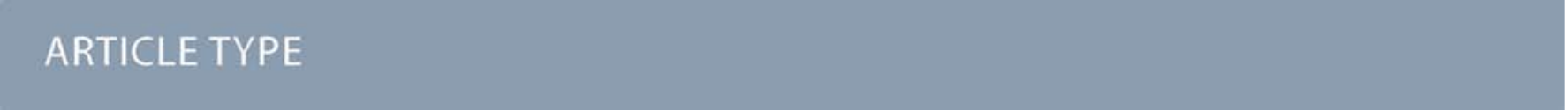}}\par
\vspace{1em}
\sffamily
\begin{tabular}{m{4.5cm} p{13.5cm} }

\includegraphics{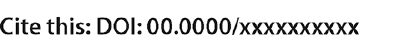} & \noindent\LARGE{\textbf{The Interfacial Structure of InP(100) in Contact with HCl and H$_2$SO$_4$ studied by Reflection Anisotropy Spectroscopy}} \\ 
\vspace{0.3cm} & \vspace{0.3cm} \\

 & \noindent\large{Mario Löw,\textit{$^{a\ddag}$} Margot Guidat,\textit{$^{a,b\ddag}$}, Jongmin Kim,{$^{a,b}$} and Matthias M.~May$^{\ast}$\textit{$^{a,b}$}} \\

\includegraphics{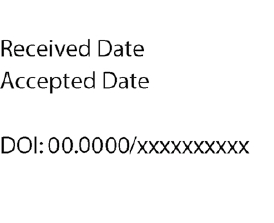} & \noindent\normalsize{Indium phosphide and derived compound semiconductors are materials often involved in high-efficiency solar water splitting due to their versatile opto-electronic properties. Surface corrosion, however, typically deteriorates the performance of photoelectrochemical solar cells based on this material class. It has been reported that (photo)electrochemical surface functionalisation protects the surface by combining etching and controlled corrosion. Nevertheless, the overall involved process is not fully understood. Therefore, access to the electrochemical interface structure under \textit{operando} conditions is crucial for a more detailed understanding. One approach for gaining structural insight is the use of \textit{operando} reflection anisotropy spectroscopy. This technique allows the time-resolved investigation of the interfacial structure while applying potentials in the electrolyte. In this study, p-doped InP(100) surfaces are cycled between anodic and cathodic potentials in two different electrolytes, hydrochloric acid and sulphuric acid. For low, 10\,mM electrolyte concentrations, we observe a reversible processes related to the reduction of a surface oxide phase in the cathodic potential range which is reformed near open-circuit potentials. Higher concentrations of 0.5\,N, however, already lead to initial surface corrosion.}

\end{tabular}



\renewcommand*\rmdefault{bch}\normalfont\upshape
\rmfamily
\section*{}
\vspace{-1cm}


\footnotetext{\textit{$^{a}$~Universität Ulm, Institute of Theoretical Chemistry, D-89081 Ulm, Germany.}}
\footnotetext{\textit{$^{b}$~Universität Tübingen, Institute of Physical and Theoretical Chemistry, D-72076 Tübingen, Germany. E-mail: matthias.may@uni-tuebingen.de}}

\footnotetext{\ddag~These authors contributed equally to this work.}

\date{}

\section{Introduction}

Renewable energies will have to play a significant role in the future energy system. Due to the intermittency of wind or solar power, however, large-scale storage of harvested energy has to be realised, for instance, by electrochemical approaches \cite{andrijanovits2012, OuldAmrouche2016}. One of the most significant scientific challenges here is understanding and controlling the electrochemical interface, where charge transfer and chemical reactions occur. The electrochemical interface of semiconductors, which are used in solar water splitting and battery applications, is a topic of intense research \cite{popovic2021, supplie2017, Smith2015, Gerischer1969, Gerischer1991}. Yet while the surface structure under \textit{operando} conditions is decisive for the performance, an atomistic picture of the interface is often not established. The last few years saw an increased effort of the community in the development and use of \textit{in situ} and \textit{operando} techniques to investigate these solid-liquid interfaces. Here, a common theme is that a complementary set of methods is typically required for the desired level of insight \cite{esposito2015, Tripathi2018, Li2020, Gourdin2021, May_coelec_photoelectrosynthetic_interfaces_2022}.

The semiconductor indium phosphide (InP) has been studied intensively for use as a photocathode for solar water splitting,\cite{Heller_InP_photocathode_1980, munoz2011} but is also involved as a major ingredient in efficient photoelectrochemical multi-junction cells \cite{May2015, Young_metamorphic_tandem_water_splitting_2017}. One of the main advantages of InP is the possibility to prepare surfaces with surface recombination velocities as low as 170\,cm/s, which leads to an efficient carrier transport across the semiconductor-electrolyte interface and reduces photovoltage losses \cite{Tournet2020, wood2014}. However, the major challenge for InP in photoelectrochemical applications is its surface corrosion and instability in aqueous solutions,\cite{Tournet2020} which then again increases surface charge-carrier recombination \cite{May_coelec_photoelectrosynthetic_interfaces_2022}.

This surface corrosion is best understood by investigating the system under operating conditions in the electrolyte. Ideally, it is then possible to design surface passivation approaches for InP \cite{munoz2011, munoz2013, lewerenz2002, schulte2002, Lebedev_InP_passivation_sodium_sulfide_2020}. Such a passivation layer should block (electro)chemical corrosion and ion transport over the solid--liquid interface while enabling efficient electron transfer. Previous investigations of InP in acidic electrolytes can be summarised as follows \cite{munoz2011, munoz2013, lewerenz2002, schulte2002, goryachev2018}: In the cathodic potential range in 0.5\,N HCl and \ce{H2SO4}, respectively, InP is reduced towards metallic indium at -0.553\,V vs.~(Ag/AgCl). In the case of HCl, the metallic In reacts further with adsorbed Cl-ions to InCl. An intermediate step that involves \ce{InCl3} is also possible, yet unlikely due to its water-solubility. Simultaneously, the phosphorous is reduced to phosphine. In the anodic regions, parts of the InCl are oxidised. However, a significant fraction remains on the surface, which results in the formation of a stable passivation layer after several electrochemical cycles that contains InCl (70 at.\%), In(PO$_3$)$_3$ (30 at.\%) and a layer of adsorbed Cl$^-$ on the surface. In \ce{H2SO4}, on the other hand, InP reacts to metallic indium and phosphine. However, the former is oxidised completely, and no stable passivation layer is observed. A possible way to obtain a stable passivation layer is by intermediate resting periods between the electrochemical cycles. \cite{goryachev2018}
For the passivation layer formed in hydrochloric acid, subsequent Rhodium deposition leads to an efficient photocathode \cite{munoz2011, munoz2013}. The treatment in an aqueous sodium sulphide solution, on the other hand, results in the removal of the native oxide and the formation of a passivation layer containing In-S and In-OH on the surface after one minute in the solution \cite{Lebedev_InP_passivation_sodium_sulfide_2020}.

Computational studies have reported on the structure and electronic properties of the pristine and oxidised InP surface using density-functional theory as well as molecular dynamics \cite{wood2014, Alvarado_water-InP_from_DFT_2022, Alvarado_InP_AlInP_oxidation_DFT_2021}. Wood et al. \cite{wood2014} investigated the interaction of gas-phase and liquid water with the pristine and oxygen-rich surfaces. Alvarado et al.\cite{Alvarado_water-InP_from_DFT_2022, Alvarado_InP_AlInP_oxidation_DFT_2021} changed water coverage adsorbed on InP from single molecules to water film, and they found the OER and HER energetics are modified due to the adsorption.

Methods used to experimentally study the formation of passivation layers on the InP in different electrolytes comprise cyclic voltammetry, X-ray photoelectron spectroscopy (XPS), and on-line electrochemical mass spectrometry (OLEMS) \cite{munoz2011, munoz2013, lewerenz2002, schulte2002, goryachev2018}. Depending on the conditions, under which surfaces are exposed to water or oxygen, the dominant pathways for surface (oxide) formation can become thermodynamically or kinetically driven \cite{Zhang_oxygen_bridges_OH-terminated_InP_2020}. Yet the question of whether these layers are well-ordered and to what extent their formation can be precisely controlled is still open. This knowledge could become instrumental for creating efficient solar water splitting cells \cite{May2015, Cheng2018}. 

In this work, we study the interface of InP(100) with hydrochloric acid (\ce{HCl}) and sulphuric acid (\ce{H2SO4}) as electrolytes. Thereby we combine cyclic voltammetry, the standard tool of electrochemistry employed in the studies discussed above,\cite{munoz2011, munoz2013, lewerenz2002, schulte2002, goryachev2018} with reflection anisotropy spectroscopy (RAS). The latter is an optical technique sensitive to changes in surface chemistry, surface states, or surface reconstruction \cite{may2014, Tereshchenko2006}. RAS enabled the investigation of adsorbates on InP \cite{may2014}, but also metallic systems were already investigated in electrolytes \cite{goletti2015, mazine2002, Sheridan2000}. We hereby show an electrolyte concentration-dependent reversibility of the cycling behaviour in \ce{HCl} and \ce{H2SO4}.

\section{Experimental}

\begin{figure}[htp]
\centering
 \includegraphics[width=10cm]{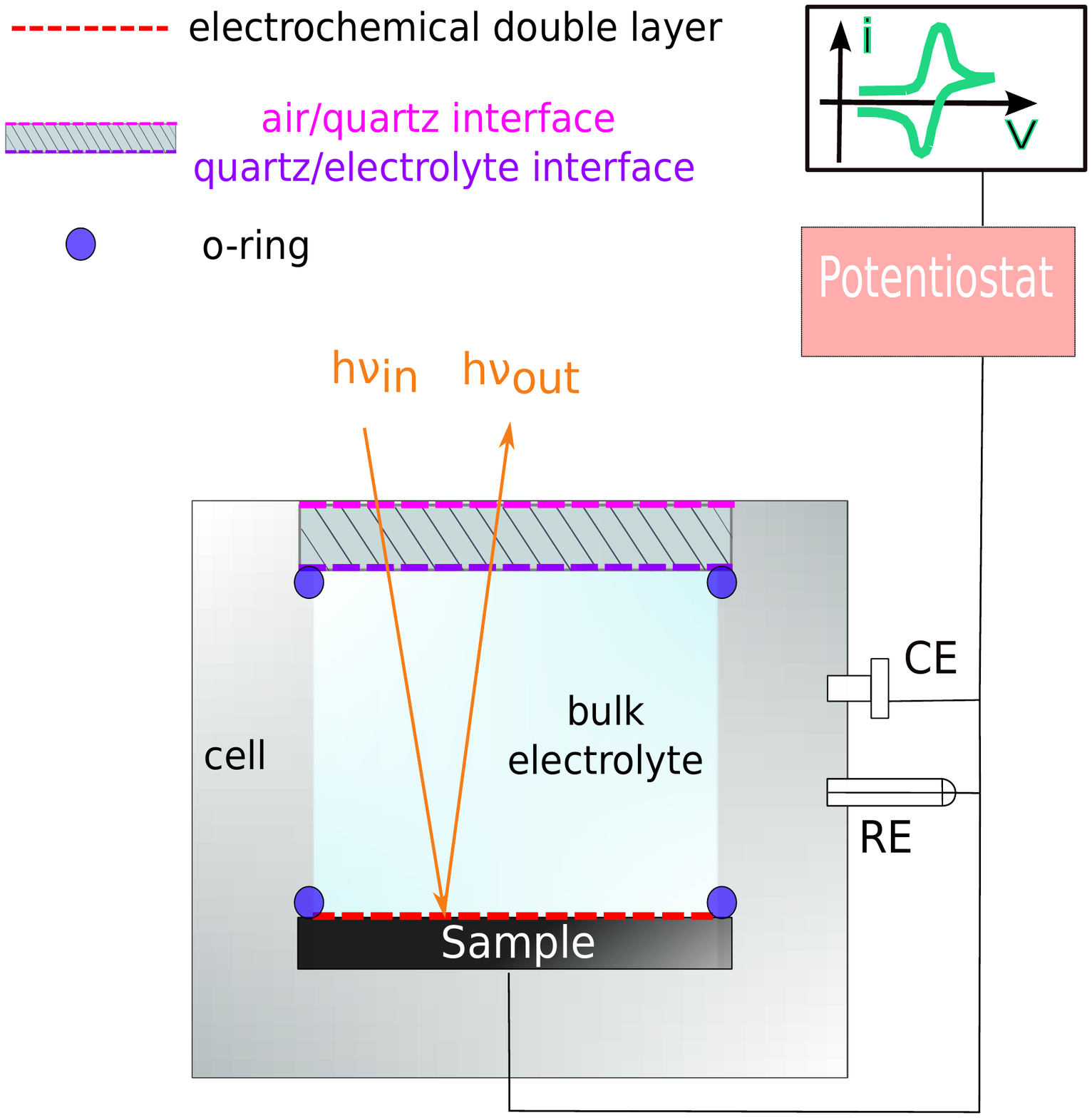}
 \caption{Cross-section of the photoelectrochemical cell set-up for electrochemical RAS measurement. Here, WE is the InP working electrode, CE is the Pt-wire counter electrode, and RE is the Ag/AgCl reference electrode.}
 \label{Cell_set-up}
\end{figure}

Samples were prepared from p-doped InP(100) (Zn-doping of $4\cdot10^{18}$\,cm$^{-3}$), single-crystalline wafers with an offcut of (0$\pm$\,0.5)\,° from CrysTec. The wafers with an epi-ready surface were cleaved and used without further pre-treatment. The samples were then mounted to a photoelectrochemical cell (PECC) from Zahner for the RAS measurement. For the electric contact, samples were back-contacted with a copper wire, fixed with silver resin (Ferro GmbH), and with some drops of GaIn eutectic (Sigma Aldrich). Pt and Ag/AgCl electrodes from Zahner were used as counter and reference electrodes, respectively. The cell was then filled with the desired electrolyte (HCl from VWR Chemicals and \ce{H2SO4} (Supelco)). During measurements, the cell was degassed with argon. For the RA spectroscopy, an EpiRAS from Laytec was employed. The electrochemical measurements were controlled with a Princeton Applied Research VersaSTAT 3F potentiostat. Fig.~\ref{Cell_set-up} shows a schematic drawing of the cell. 

 Reflection anisotropy spectroscopy was used here as a spectroelectrochemical method. RAS is a highly surface-sensitive tool that can investigate changes in surface chemistry, surface states, and surface reconstruction with a near-normal-incidence reflection set-up \cite{Haberland_RAS_setup_1999}. In this method, linearly polarised light impinges at near-normal incidence on the surface. The difference in the reflectivity, $\Delta r$, with respect to two orthogonal directions in the surface plane ($x, y$) is detected by analysis of the polarisation as a function of the photon energy, scaled with the overall reflectivity, $r$:
 
 \begin{equation}
\centering
\mbox{RAS}:=\frac{	\Delta r}{r} = \frac{2(r_x - r_y)}{r_x + r_y}; r\in\mathbb{C}
\label{eqRAS}
\end{equation}

In the case of a crystal, where the bulk is optically isotropic -- as for the (100) surfaces of zincblende lattices here -- there is no contribution to the signal from the bulk, but only from the (near) surface. Then, RAS becomes highly surface sensitive. This is a significant advantage to most other spectroscopies working with photons as those typically  also probe a significant depth of the bulk material \cite{may2014, weightman2005, martin2000}. A baseline correction is applied to all the spectra using the signal of an optically isotropic Si(100) crystal in water. Cyclic voltammetry measurements were performed between -2\,V and +1.5\,V (vs. Ag/AgCl) with a scan rate of 0.05\,V/s, starting from open circuit potential (OCP) towards the negative terminus for three cycles for low concatenation (0.01\,M) electrolyte, respectively. For higher-concentration electrolytes (0.5\,N), the cyclic voltammogram (CV) was performed between -0.75\,V and +0.45\,V (vs.~Ag/AgCl) with a scan rate of 0.05\,V/s, starting from the OCP towards the negative terminus for ten cycles, respectively.

Further on, potentiostatic measurements were performed at the potentials of -2\,V, -1\,V, 0\,V, and +1.5\,V vs. Ag/AgCl, respectively. Potential step experiments were performed to investigate the stability of the surface at cathodic potentials. In this case, the potential was set to OCP for 10\,s, then to -1\,V for 30\,s, then to OCP for 30\,s, to -2\,V for 30\,s, and finally to OCP for 20\,s. This cycle was then repeated two more times. During the electrochemical experiments, RA spectra were recorded continuously, or the anisotropy was measured time-resolved at fixed photon energy (transient mode).

For the computational spectroscopy, density-functional theory (DFT) calculations using the CP2K code \cite{Kuehne_2020} were first performed to obtain the relaxed structure of the slabs. The slab consists of 17 layers. To avoid spurious interactions between replica of the slab under 3D-periodic boundary conditions, we set a vacuum spacing of 20\,\AA{} between the slabs. Exchange and correlation effects were described by means of the generalised gradient approximation (GGA) using Perdew-Burke-Ernzerhof (PBE) parameterisation \cite{Perdew_1996}. The plane-wave cutoff was set to 800\,Ry. We used Goedecker–Teter–Hutter (GTH) pseudopotentials to represent core electrons and valence electrons. After the structural optimisation, the atomic coordinates were transferred to the plane wave/pseudopotential code QUANTUM ESPRESSO \cite{Giannozzi_2009}. For these calculations, optimised norm-conserving Vanderbilt pseudopotentials were adopted \cite{schlipf_2015}. The Brillouin zone (BZ) was sampled on a 10$\times$10$\times$1~$\mathrm{\mathbf{k}}$-grid with an energy cutoff of 60\,Ry. The RAS calculations were performed using the Yambo code \cite{Marini_2009, Sangalli_2019}. We employed the RPA-IP approach for the microscopic dielectric function. To correct the underestimated bandgap calculated by PBE, a scissor shift was adopted. In the RAS calculations, we used an experimental bulk dielectric function \cite{Aspnes_1983}.

\section{Results \& Discussion}

\begin{figure}[h!]

 \centering
 \includegraphics[width=17.1cm]{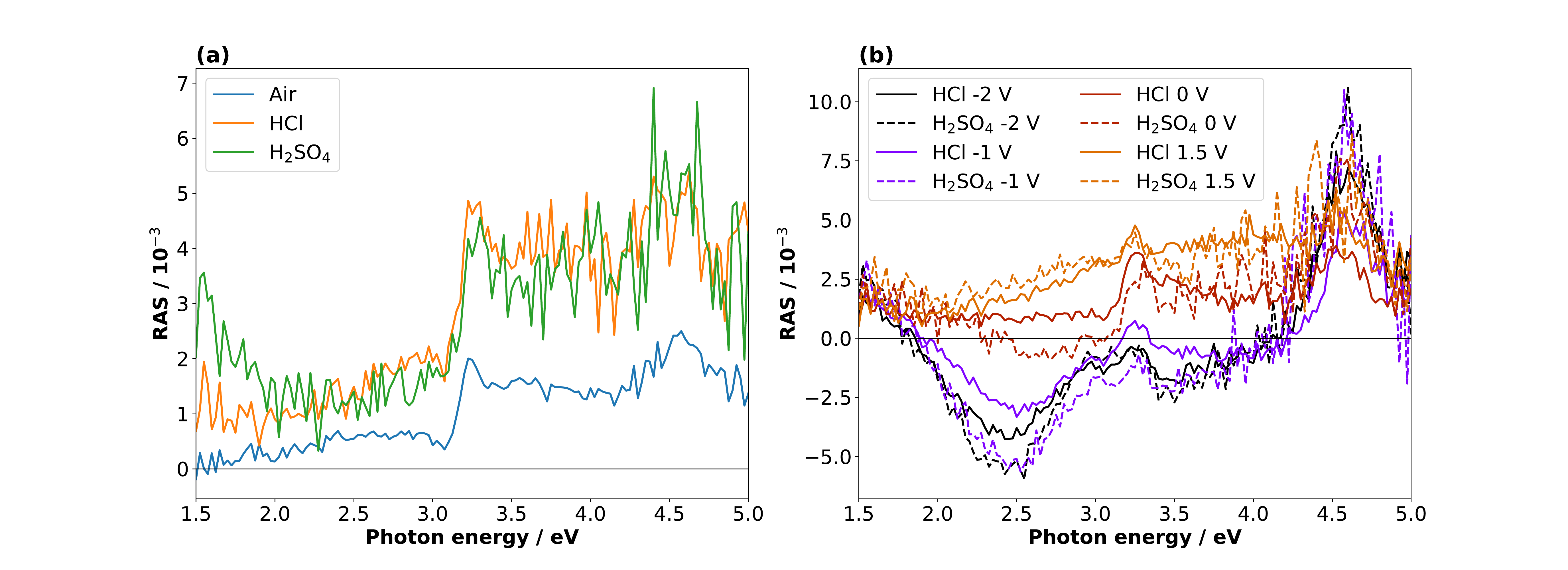}
 \caption{(a) RA spectra of epi-ready InP(100) in air and immersed in 0.01\,M HCl and \ce{H2SO4} at open-circuit potential. (b) RA spectra at -2\,V, -1 \,V, 0\,V and 1.5\,V of the InP(100) surface in 0.01\,M \ce{HCl} (solid line) and 0.01\,M \ce{H2SO4} (dashed line) solution.} 
 \label{RAS_Spectra}
\end{figure}

Fig.~\ref{RAS_Spectra}(a) shows the InP(100) spectra in 0.01\,M hydrochloric and 0.01\,M sulphuric acid, respectively, at OCP (around 0.2\,V vs.~Ag/AgCl) as well as the epi-ready crystal in air under ambient conditions without any pre-treatment. All three spectra show two positive peaks at 3.25\,eV and 4.55\,eV due to surface-modified bulk transitions \cite{may2014}. The presence of electrolytes only slightly influences the RA spectra with an upwards shift in the RAS baseline, which can also arise from imperfect baseline correction. The noise in the high-energy region of the spectra is due to a reduced overall reflectivity from absorption and reflection losses in the electrolyte and at the air--electrolyte interface, respectively.

\begin{figure}[ht]

 \centering
 \includegraphics[width=17.1cm]{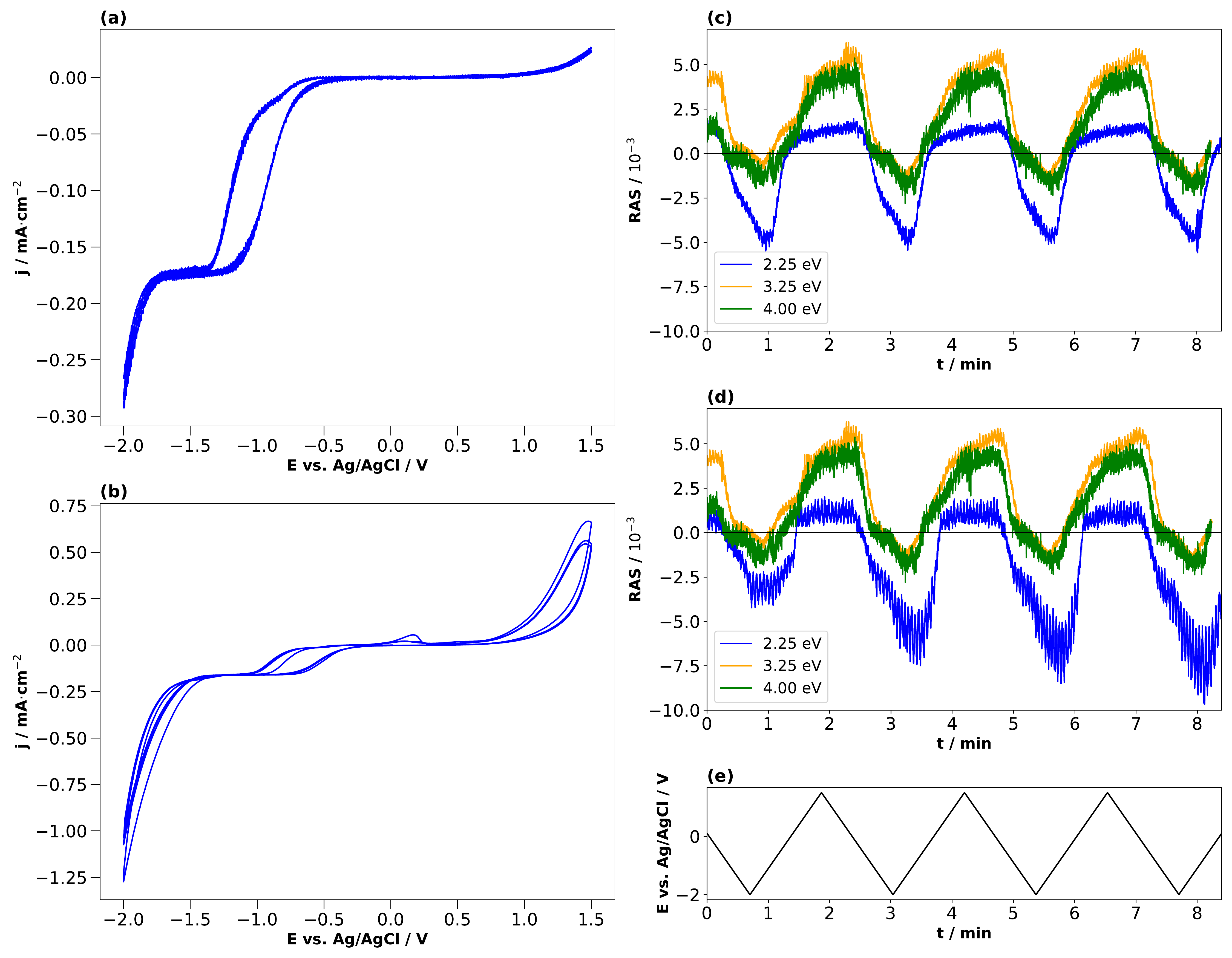}
 \caption{Cyclic Voltammograms of InP in 0.01\,M (a) HCl with associated transients (c), and in 0.01\,M (b) H$_2$SO$_4$ with associated transients (d). (e) Potential plotted against Ag/AgCl. Scan from OCP towards -2\,V and than to 1.5\,V for 3 cycles, with a scan rate of 0.05\,Vs$^{-1}$.}
 \label{CV_Trans_low}
\end{figure}

In the next step, different potentials were applied to investigate their effect on the optical anisotropy and hence the interfacial structure. To this end, we performed potentiostatic measurements at 0\,V, -1\,V, and -2\,V in the cathodic range and +1.5\,V in the anodic potential range vs.~Ag/AgCl, respectively, during the acquisition of RA spectra. The RA spectra in Fig.~\ref{RAS_Spectra}(b) at anodic potentials shift towards more positive anisotropies. In contrast, RA spectra taken at cathodic potentials develop a negative peak for photon energies below 3\,eV and the rest of the spectrum shifts to slightly more negative anisotropies. Additionally, the peak at 4.55\,V is much more intensive than at 0\,V or OCP. These spectra suggest an ordered surface structure in the cathodic and anodic potential ranges, but with different surface structure or composition. The similarity of the spectra at OCP or 0\,V vs. Ag/AgCl with the epi-ready oxide indicate a similarity of the interface structure under these conditions.

To investigate the dynamic behaviour of the interface between these states during the cycling in the electrolyte, we changed the potential continuously between two terminals. The CV in 0.01\,M HCl is shown in Fig~\ref{CV_Trans_low}(a). In the cathodic region, the cathodic current density increases from around -0.4\,V, reaches a plateau at -1.3\,V, and increases further after -1.7\,V. This current plateau, observed also for \ce{H2SO4}, corresponds to the photocurrent from the white-light RAS measurement spot. Notable is a small deflection around -0.85\,V that occurs only in the forward scan from OCP towards cathodic potentials. In the anodic region, there is an increase in the current density at potentials greater than 1\,V. The first cycle shows the highest current density.

In the CV in 0.01\,M \ce{H2SO4} in Fig.~\ref{CV_Trans_low}(b) the cathodic current density increases beyond -0.5\,V, reaches a plateau at -1.1\,V and increases further beyond -1.4\,V. In this case, however, the second step after -1.4\,V is steeper than for HCl and simultaneously leads to a higher current density value at the potential terminus, where the current is dominated by the hydrogen evolution reaction \cite{goryachev2018}. For the first forward scan towards cathodic potentials, the cathodic current onset starts at -0.7\,V, but at -0.8\,V for the subsequent scans. Also here, a small deflection of the cathodic current is observed around -0.6\,V. This is in the potential region where, albeit at higher electrolyte concentrations, the reduction of InP towards metallic indium and phosphine was reported at around -0.55\,V vs. Ag/AgCl \cite{lewerenz2002, schulte2002, munoz2011, munoz2013}. The cathodic current density at the potential terminus decreases from the first cycle on. Above +1\,V, we observe an anodic current, which decreases over the three cycles. In the first cycle, a small oxidation wave can be observed at around +0.3\,V, which is absent for the CV taken with HCl. According to the literature, this anodic peak could due to the formation of indium phosphate \cite{muthuvel2005}.

During the CV measurements, transients were recorded \textit{operando} at the energies 2.25\,eV, 3.25\,eV, and 4.00\,eV. The photon energies were chosen since changes in the spectra at different potentials in Fig.~\ref{RAS_Spectra}(b) are most pronounced here. These transient measurements enabled us to observe the potential-induced changes on the surface during cycling at higher time-resolution and evaluate the reversibility with respect to the interfacial structure. Fig.~\ref{CV_Trans_low} shows the periodic changes of the anisotropy in these transients. The minimum of the optical anisotropy occurs under the application of a cathodic potential. In contrast, the maximum is associated with the anodic region. Shape and amplitude of the transients remain virtually constant for HCl, with the exception of the transients at 3.25 and 4\,eV, which become more negative from the first cathodic half-scan to the subsequent ones. The case of \ce{H2SO4} shows distinct differences. Here, the negative extremum of the RA amplitude at 2.25\,eV increases from cycle to cycle, yet not for the higher photon energies, which, similar to HCl, only show a slightly less negative amplitude in the first cathodic scan. From these different signal shapes at different potentials and electrolytes, we can try to deduce the reactions on the interface of InP with the respective electrolyte.

At the beginning of the experiment, the epi-ready InP(100) wafer is covered by a thin, well-defined oxide layer. In the presence of an acidic electrolyte, the oxide layer is expected to dissolve when scanning the potential from OCP towards the cathodic region \cite{munoz2011, munoz2013, lewerenz2002, schulte2002, goryachev2018}. The well-defined, reproducible RA spectrum that we observe at cathodic potentials indicates that this oxide-free layer is well-ordered. Yet the spectral shape does not clearly correspond to one of the main characteristic surface reconstructions of InP(100), the 'P-rich' $p(2 \times 2)/c(4 \times 2)$ reconstruction or the mixed-dimer 'In-rich' $(2 \times 4)$ reconstruction. Both these surfaces exhibit a pronounced negative anisotropy around 1.8-2\,eV, associated with surface dimers \cite{may2014}, while here, the peak observed here is located at 2.5\,eV. The strong positive anisotropy around 4.5\,eV, however, is a common feature with the In-rich surface \cite{may2014, Tereshchenko2006}. Scanning back to OCP or 0\,V vs.~Ag/AgCl, the spectrum is again similar to the epi-ready oxide. This suggests a re-formation of the oxide phase in this potential region for both electrolytes. In both electrolytes, the negative anisotropies are slightly reduced for the first cycle, which could suggest that the re-formed oxide layer is of slightly different nature, possibly with a higher degree of ordering as indicated by the higher transient amplitudes. The observation that both CV and RAS transients remain then constant for HCl indicates a highly reversible process. This would be compatible with the adsorption of chloride on an In-terminated surface, leading to the formation of an InCl surface film starting from -0.55\,V\,(Ag/AgCl) \cite{lewerenz2002}, which would also explain the differences at lower photon energies to \ce{H2SO4}. The further change of the anisotropy at cathodic potentials beyond -1.5\,V could then be related to the reordering of the (near-)interface water \cite{Alvarado_water-InP_from_DFT_2022} or the adsorption of protons/hydrogen on the surface.

In \ce{H2SO4}, only a fraction of the indium is expected to react with the sulphate ions because of sulphate's weaker interaction with InP compared to the chloride-InP interaction in HCl \cite{munoz2013, schulte2002, Lewerenz_1984}. Therefore, metallic indium and \ce{In_{x}(SO4)_{y}} species would be present at the interface. According to the literature, the metallic Indium and InCl compounds are partially oxidised to \ce{In^{3+}} in the anodic potential range, forming Indium oxide, and Indium phosphate at high electrolyte concentrations \cite{lewerenz2002, schulte2002}. Interestingly, the increasing amplitude of the negative anisotropy with ongoing cycling in the \ce{H2SO4} electrolyte is associated with a higher overpotential for the hydrogen evolution reaction.

\begin{figure}[ht]

 \centering
 \includegraphics[width=17.1cm]{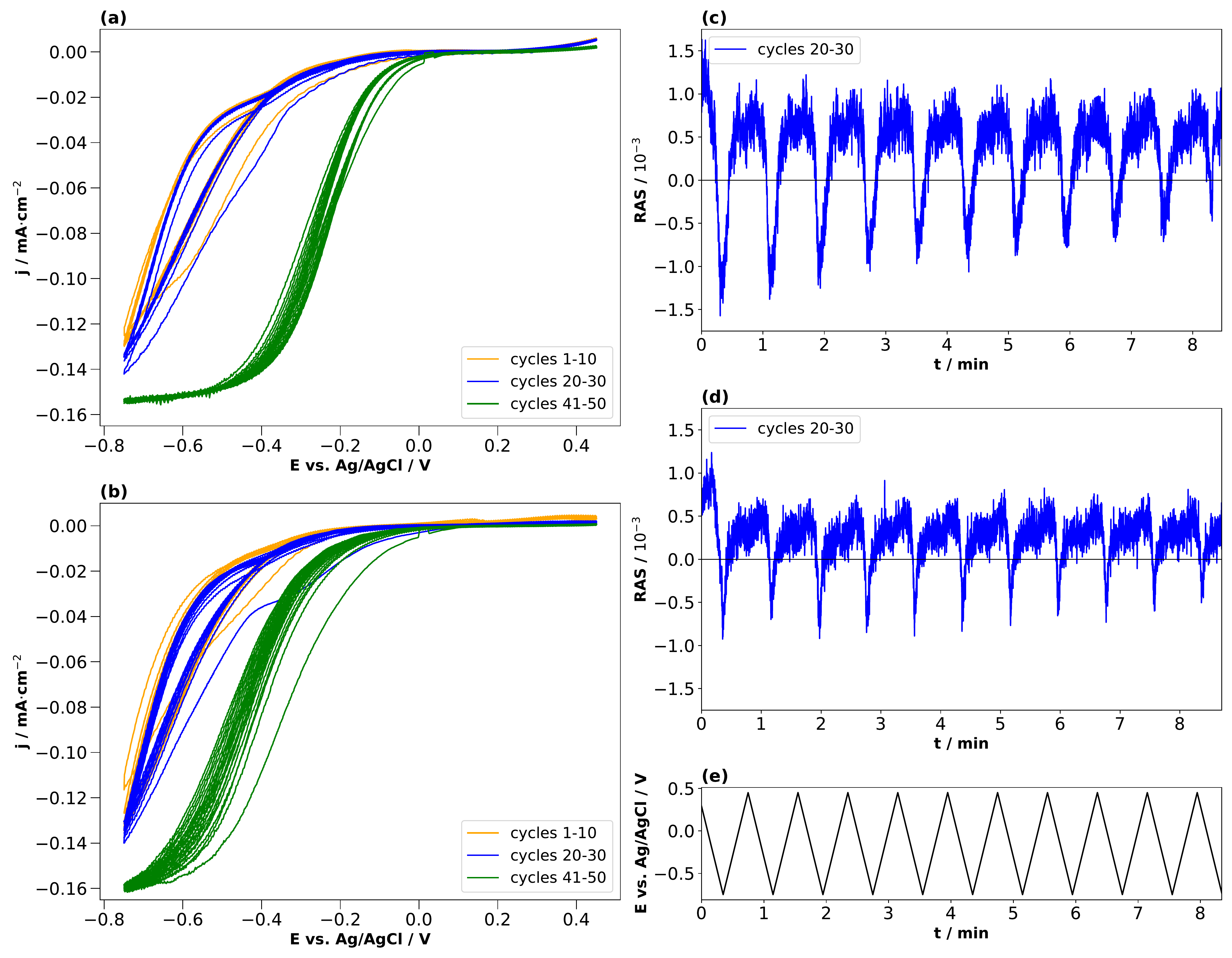}
 \caption{Cyclic voltammograms of InP a) HCl (0.5\,M) and b) \ce{H2SO4} (0.25\,M) with corresponding transients at 2.25\,eV for cycles 20-30 (c,d) and the applied potentials (e). Potential plotted against Ag/AgCl. Scan from open-circuit potential towards -0.75\,V and then to +0.45\,V with a scan rate of 0.05\,V/s. The orange curve shows the cycles 1-10, the blue curve shows the cycles 20-30, , and the green curve shows the cycles 41-50.}
 \label{CV_Trans_high}
\end{figure}

In the literature, conditioning of InP for use as a photocathode was performed typically at higher electrolyte concentrations of 0.5 or 1\,N, and these were also the concentrations used to study the resulting surface composition by photoelectron spectroscopy \cite{lewerenz2002, schulte2002}. We therefore also applied the same approach to these higher electrolyte concentrations. Figs.~\ref{CV_Trans_high}(a) and (b) show the resulting cyclic voltammograms for both HCl and \ce{H2SO4}. The CV changes its shape from cycle to cycle, effectively reducing the overpotential for the hydrogen evolution reaction, as observed similarly in the literature \cite{schulte2002}. 

Unlike for the lower concentrations, the amplitudes of the transients at 2.25\,eV decrease over the cycles in the cathodic potential region. This occurs continuously for HCl, less so for \ce{H2SO4}. This hints towards a decreased interface ordering with ongoing cycles, consistent with inline XPS studies suggesting the formation of a layer of mixed oxides with a thickness in the order of a single or few nanometres \cite{lewerenz2002, schulte2002,munoz2013}. Consequently, cycling in higher electrolyte concentrations induces slightly different surface reactions reducing the degree of ordering of the interface. A comparison with the literature on InP(100) in 0.5\,M HCl \cite{lewerenz2002, schulte2002} shows that the build-up of a passivation layer progressively transforms the surface into a relatively thick, amorphous layer. This lack of reversibility is observed through the evolution of the negative anisotropy in Figs.~\ref{CV_Trans_high}(c,d), decreasing from cycle to cycle. In HCl, the passivation layer is composed of InCl, indium phosphate and indium oxide \cite{lewerenz2002, schulte2002}.  In contrast, the passivation layer from cycling in \ce{H2SO4} consists of metallic In, \ce{In_{x}(SO4)_{y}}, indium phosphate and indium oxide.

\begin{figure}[ht]

 \centering
 \includegraphics[width=8.3 cm]{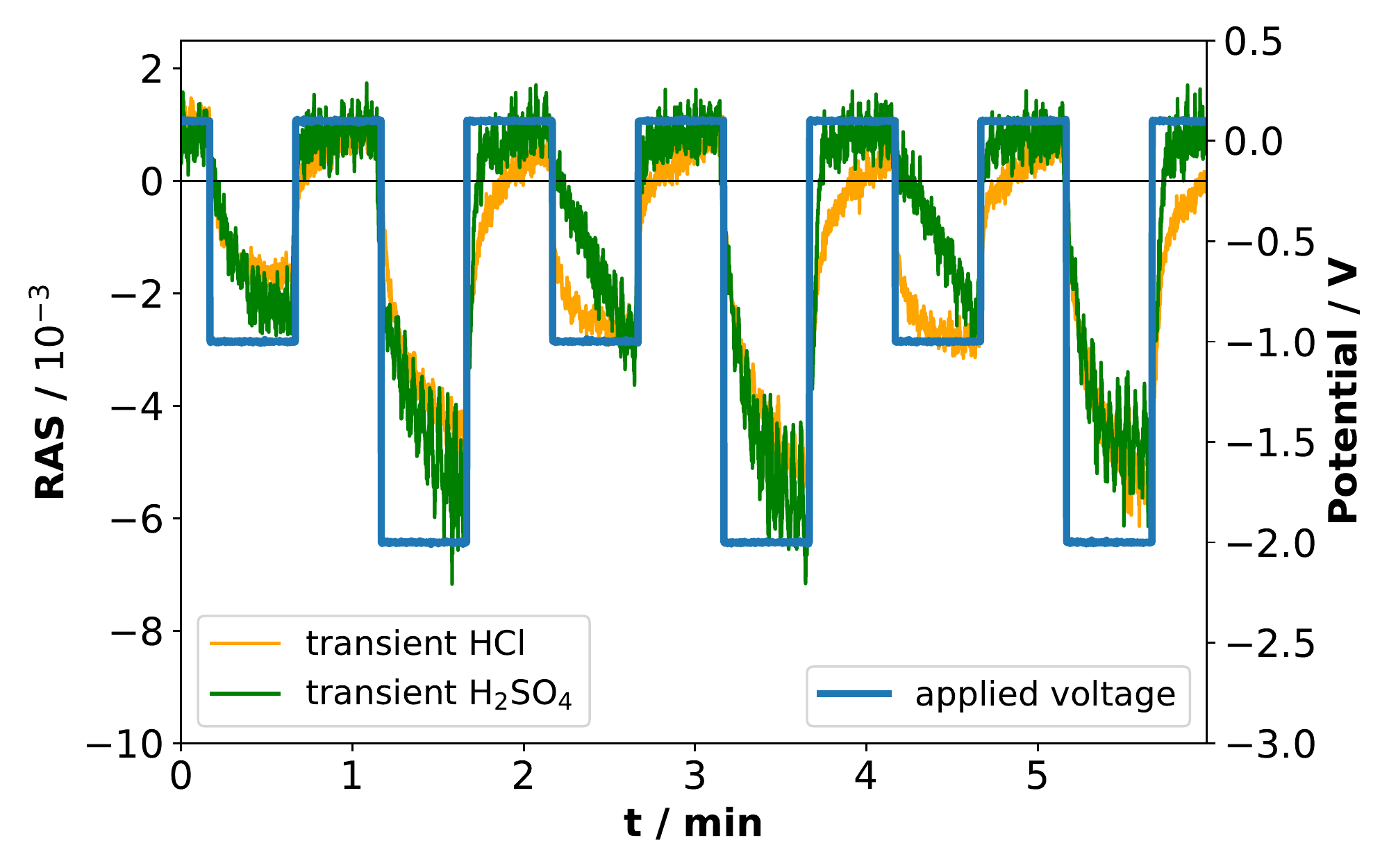}
 \caption{Potential-step and transient measurements at 2.25\,eV in 0.01\,M HCl and in 0.01\,M \ce{H2SO4}.}
 \label{Pot_step}
\end{figure}

For a more detailed insight, information on the exact surface composition would be instrumental, especially for the structurally well-ordered interface in lower electrolyte concentrations. Alas, in virtually all cases, RAS does not allow to directly derive composition and exact structure from experimental spectra alone, requiring either correlation with other techniques or computational spectroscopy \cite{May_water_adsorption_calc_RAS_2018}. A straightforward choice for probing the surface/interface composition is X-ray photoelectron spectroscopy. Standard XPS does, however, require samples prepared at the desired potential to be transferred from the electrolyte to ultra-high vacuum (UHV). Especially for ultra-thin interface layers, it is important to understand to what extent the surface created at a certain potential will be stable in the absence of applied potential and/or electrolyte. Therefore, we tested the stability of the newly formed interface at cathodic potentials through potential-step experiments during which transients were recorded. This experiment aims to assess the kinetics of formation and disintegration of the interface formed at -2\,V and -1\,V, respectively. Depending on how fast these changes occur, one can try to assess whether the stability window of the cathodically formed interface is large enough for transfer to UHV. Fig.~\ref{Pot_step} shows the resulting transients in both electrolyte at 2.25\,eV.
 
When the potential is applied, the RAS amplitudes of the transient spectrum decrease towards its negative extremum in equilibrium, but with different time-constants. Especially for HCl at -1\,V, there appears to be a fast process in addition to a slower signal decrease, while the transient in \ce{H2SO4} decreases more slowly. After setting the potential back to OCP, the anisotropy in \ce{H2SO4} changes back to the original value at OCP, while the process is slower in HCl. One of the two processes observed in HCl could be related to the formation and dissolution of the above-mentioned InCl layer, which is not present for the \ce{H2SO4} case. Yet the time-constants clearly indicate that the time-window for a transfer to XPS is very small and conclusive results would very likely require ambient-pressure XPS similar to Ref.~\cite{Zhang_oxygen_bridges_OH-terminated_InP_2020}, but at different applied potentials.

\begin{figure}[ht]
\begin{center}
\includegraphics[width=1.0\linewidth]{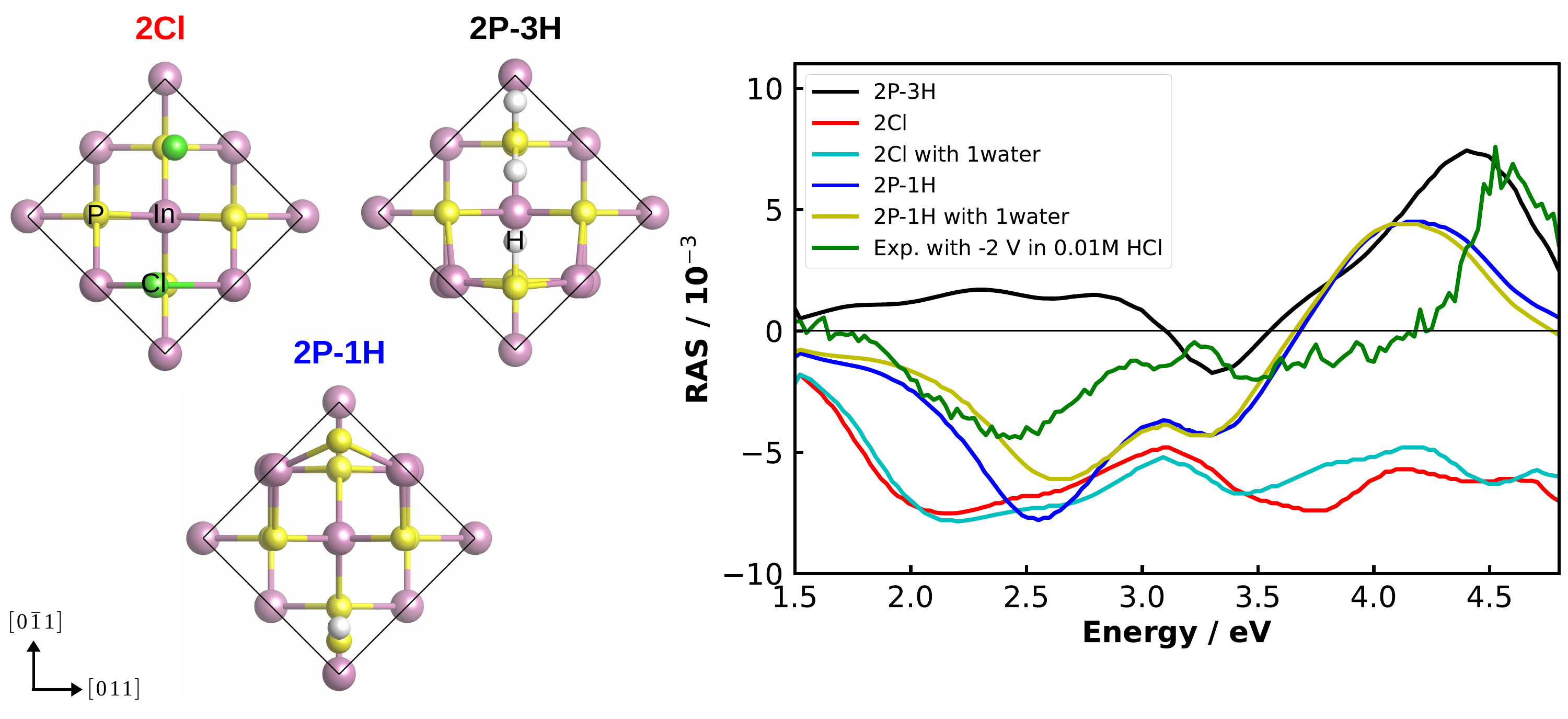}
\end{center}
\caption{Left: Top view of the fully Cl-covered surface, as well as the surface with one and three adsorbed hydrogen atoms of InP(100) (2Cl, 2P-1H, 2P-3H, respectively). The black line indicates the surface unit cell. Right: Computed and experimental (green) RAS of the InP(100) in contact with 0.01\,M HCl solution under the applied potential of -2\,V.}
\label{Comp_fig}
\end{figure}

As an alternative to correlation with another experimental method for a more detailed understanding of our experimental RA spectra, we performed computational RAS on structures that could be expected to be relevant for InP(100) in contact with an electrolyte. In Fig.~\ref{Comp_fig}, we compare the experimental RA spectrum of InP(100) in 0.01\,M HCl at -2\,V with calculated spectra. For the latter, we model three systems in which phosphorus atoms at the InP surface are fully substituted with chlorine, forming an InCl layer (termed 2Cl), and one or three hydrogen atoms are adsorbed on the surface (2P-1H and 2P-3H). We further consider adsorption of one water molecule on the 2Cl and 2P-1H structures, as water molecules can significantly alter the RAS signature \cite{May_water_adsorption_calc_RAS_2018}. When comparing computational to experimental RA spectra, it has to be noted that experimental surfaces would be expected to be not as 'perfect' as the computational structure, where all of the surface is in the same state. This would translate to lower intensities for the experimental spectra. We find that elements of the computed spectra for the 2Cl and 2P-1H, yet also for the 2P-3H surfaces resemble the experimental signature. The calculation for the 2Cl surface reveals a constantly negative anisotropy with negative peaks around 2.2 and 3.8\,eV and a shoulder around 2.7\,eV, the additional water molecule slightly modifies the spectrum, but the general shape is conserved. A noticeable difference to the experimental RA-signal is that the positive anisotropy at 4.5\,eV does not appear in the calculated spectrum. For the case of 2P-1H, two pronounced negative anisotropies around 2.5\,eV and 3.3\,eV are observed. Unlike the 2Cl surface, the strong positive feature, which also arises in the experimental spectrum, is shown at 4.2~eV. Compared to the experiment, this peak is red-shifted by around 0.35\,eV, while the minimum of the anisotropy is slightly blue-shifted and the shift increases with adsorbed water. The 2P-3H structure, on the other hand, displays a rather flat positive anisotropy up to 3\,eV, a slight negative peak at 3.3\,eV, and a strong positive anisotropy at around 4.4\,eV. The latter feature is very similar to the high-energy peak of the experimental spectrum. In general, single water molecules doe not play a crucial role for these spectra. Minor discrepancies between the experiment and calculations in terms of the energies would be expected from bandgap errors of the exchange-correlation functional and neglecting defects, oxidation, and domain boundaries on the surface. These initial results suggest a coexistence of H-terminated and Cl-terminated surface, which would lead to a superposition of the spectra of the respective structures. While the computational results give already a first hint towards the real surface structure, a surface phase diagram followed by computational spectroscopy of the -- numerous -- different structures is required to probe a higher fraction of the chemical space. In addition, multiple water layers and applied potentials might be required \cite{May_water_adsorption_calc_RAS_2018}. Due to the significant involved computational effort, however, this will have to be left to future work.

To summarise, our results indicate different mechanisms for the two electrolyte concentrations. The transients measurements of InP(100) in low acidic concentrations (0.01\,M, Fig.~\ref{CV_Trans_low}) reveal persistently reversible changes in the surface ordering during cycling. Starting from open-circuit potential, a well-ordered, ultra-thin oxide layer appears to be dissolved and, in the case of HCl, replaced by a monolayer of InCl. The spectra found in the cathodic potential range show a strong negative anisotropy around 2.5\,eV for both HCl and \ce{H2SO4}, yet with slight differences. Hydrogen adsorption appears to be involved as well for both electrolytes. Setting the potential back to open-circuit, this oxide layer is re-formed and appears to exhibit a higher degree of ordering than the original, epi-ready oxide on the wafer as indicated by a slightly more pronounced spectrum after the first cycle. Also in the anodic potential range, a spectrum different from OCP is found, but less distinct, which could indicate a lower degree of ordering in this potential range. The reversibility indicates, that potential cycling does not involve further etching of the surface, which is typically accompanied by a gradual loss of interface ordering. Potential cycling in higher electrolyte concentrations of 0.5\,N, on the other hand, leads to a gradual change in the shape of the cyclic voltammograms, associated with a gradual loss of interface ordering, when the nm-thick mixture of oxides and phosphates \cite{lewerenz2002, schulte2002, muthuvel2005} is formed.

\section{Conclusions}

Using \textit{operando} electrochemical reflection anisotropy spectroscopy for the study of electrochemical InP(100) interfaces in contact with HCl and \ce{H2SO4}, we find the potential-dependent formation of well-ordered InP--electrolyte interfaces for low electrolyte concentrations. Transient measurements indicate for both electrolytes that there is a reversible formation and removal of ultra-thin oxide layer, yet only in the case of low-concentration (0.01\,M) electrolytes. In the case of higher-concentration electrolytes the interface gradually transitions into an unordered state.

Time-resolved potential step measurements show that the surface prepared at cathodic potentials is not stable when the potential is set back to OCP, which hampers the study of the exact chemical nature of this interface state by standard X-ray photoelectron spectroscopy. While initial results from computational RA spectroscopy corroborate the hypothesis that the structures formed in HCl under cathodic potentials involve InCl and adsorbed hydrogen, further computational studies, also under applied potentials, will have to be performed for a more definitive insight.

In summary, our study shows the potential of electrochemical reflection anisotropy spectroscopy for the preparation of well-structured interfaces in the electrolyte. These could then serve as a starting point for subsequent (photo)-electrochemical functionalisation, such as deposition of passivation layers and catalysts, to be used for solar-water-splitting devices \cite{May2015}. As the process observed here for InP resembles the formation of of solid--electrolyte interphases (SEI) \cite{May_coelec_photoelectrosynthetic_interfaces_2022}, electrochemical RAS also has potential as a technique to gain a microscopic, time-resolved insight into SEI processes on battery electrodes.

\section*{Conflicts of interest}
There are no conflicts to declare.

\section*{Acknowledgements}

This work was funded by the Deutsche Forschungsgemeinschaft (DFG, German Research Foundation) under Germany's Excellence Strategy -- EXC 2154 -- project number 390874152 as well as DFG project number 434023472. We acknowledge the state of Baden-Württemberg, through bwHPC, and DFG, through grant no.~INST 40/575-1 FUGG (JUSTUS 2 cluster) for computational resources.



\providecommand*{\mcitethebibliography}{\thebibliography}
\csname @ifundefined\endcsname{endmcitethebibliography}
{\let\endmcitethebibliography\endthebibliography}{}

\end{document}